# Times in noise-based logic: increased dimensions of logic hyperspace

LASZLO B. KISH

[1] Texas A&M University, Department of Electrical and Computer Engineering, College Station, TX 77843-3128, USA

**Abstract.** Time shifts beyond the correlation time of the logic and reference signals create new elements that are orthogonal to the original components. This fact can be utilized to increase the number of dimensions of the logic space while keeping the number of reference noises fixed. Using just a single noise and time shifts can realize exponentially large hyperspaces with large numbers of dimensions. Other, independent applications of time shifts include holographic noise-based logic systems and changing commutative operations into non-commuting ones. For the sake of simplicity, these ideas are illustrated by deterministic time shifts, even though random timing and random time shifts would yield the most robust systems.

## 1. Introduction

Noise-based logic (NBL) [1-11] has been introduced to explore the possibility to use stochastic processes (noises) and their products/superpositions as carriers of the logic information, similarly to the stochastic neural spike sequences in the brain [3,4]. For the identification of the logic signals, a reference noise systems of $N$ "noise-bits", consisting of $2N$ orthogonal noises have been used (two noises represented the two bit values) [1]. Depending on how the incoming noise signals are analyzed in the NBL gates, the systems explored were either correlator-based [1] or instantaneous [5,6] NBLs, where the stochastic spike sequence based brain logic belongs to the latter class. Correlator-based solutions are slow but robust against external and internal noises while instantaneous versions are fast but less resilient. Product strings of noise-bits [1,7] and their superpositions represent a hyperspace with $2^N$ dimensions corresponding to the expressivity of $2^N$ parallel classical bits in a single wire [7], similarly to qubits. That offers an exponential increase of equivalent classical bit size in representing multivalued logic spaces and a similar speed increase in certain special-purpose operation at only a polynomial increase of the requirements for hardware and time complexity. The superposition of the first $2^N$ integer numbers [7] and all the single-bit quantum gates can efficiently be emulated [8] by instantaneous NBL with an exponential $O(2^N)$ speedup [8-10] compared to classical binary computers. However the implementation of two-bit operations, particularly the CNOT gate, on these superpositions are yet awaiting solution before Shor algorithm can efficiently implemented [8].

In this short note, we introduce the application of longer-then-the-correlation-time (LTC) time shift of the noise-bit signals to generate new noise bits. In an NBL system with $N$ noise-bits, with the application of $M$ steps of LTC time shifts, the number of classical bits represented by a single wire is $2^{NM}$.

## 2. Time offers new dimensions in NBL

Suppose, we have a single noise source with instantaneous amplitude $U(t)$ and correlation time $\tau_0$. The LTC time shifted noise is defined as:

$$U(t+\tau) \qquad (1)$$

where $\tau \geq \tau_0$ so that the crosscorrelation between the original and time-shifted noises satisfies

$$\langle U(t)U(t+\tau)\rangle = 0 . \qquad (2)$$

To have a binary NBL with a single noise-bit, for the 0 and 1 logic values represented by the orthogonal noises $V_0(t)$ and $V_1(t)$, where $\langle V_0(t)V_1(t)\rangle = 0$, one can choose:

$$V_0(t) = U(t) \quad \text{and} \quad V_1(t) = U(t+\tau) \qquad (3)$$

thus the two orthogonal noises necessary for a non-squeezed binary NBL system can be realized by a single noise generator and a single deterministic time shifter.

Similarly, the $2N$ orthogonal noise needed for an $N$ noise-bit system can be realized from a single noise source and $2N-1$ deterministic time shifts:

$$V_{1,0}(t) = U(t), V_{1,1}(t) = U(t+\tau), \dots ,$$
$$\dots , V_{N,0}(t) = U[t+(2N-2)\tau], V_{N,1}(t) = U[t+(2N-1)\tau] \qquad (4)$$

Similarly, if originally we had the $2N$ orthogonal noise needed for an $N$ noise-bit system and wanted to expand the system, by executing a single deterministic time shift on all the $2N$ noises will double the effective noise bits of the system resulting in the representation of $2^{2N}$ classical bits in a single wire.

In general, in an NBL system with $N$ noise-bits, with the application of $M = 2kN$ steps of LTC time shifts, where $k$ is an integer number, the number of classical bits represented by a single wire is $2^{N+M/2}$. Accordingly, the number of dimensions of the logic hyperspace increases by a factor of $2^{M/2}$.

## 3. Some applications

One of the applications is the one described above: increasing the effective number of noise-bits by time shift operations.

Another application is to create a "holographic" superposition where applying various time shifts on the superposition (or that of on the reference system) yields new interpretations of the logic information. For example, at a certain time shift, a certain product string may be present in the superposition and at a different time shift it may show up as another bit string.

Yet another application is to make originally commutative gate operations to be non-commutative by executing single time shifts during the operation (and proper time shift alternatives of the reference system).

## 4. Example

Using a binary random telegraph wave (RTW) [5,6,11] with deterministic time steps results in a square with fixed periods $T$ and +1 or -1 amplitudes (each with probability of 0.5) over the different periods. The correlation time of this process is $T$ and the autocorrelation function has an abrupt cutoff at $T$. Thus a single deterministic time shift with of $T$ results in a new RTW, which has zero crosscorrelation with the original one. To generate $N$ noise bits from a single RTW, a shift register with $2N$-1 elements is needed.

## 5. Deterministic versus random time shift

What is the price we pay for deterministic time shifts? Losing some of the resilience of the noise bits and their values. Originally the noises representing each noise-bit values were "fingerprints". The noise bits and their values could be transformed into each other by a simple transformation. This claim does not hold for the time shifted noise-bits and their values: a simple deterministic transformation, and inverse time shift restores the original bit value.

This is a point where random time shifts are more advantageous. They can conserve part or the whole original resilience. Continuum noise based bit values can have a fixed random time shift. The original noise can still be restored by the same inverse time shift however the restoration of each noise-bit value will need inverse transformations by the actual time shift, thus a single time shift value is not enough. In a random-period RTW and the random spike sequence system (brain logic), randomly chosen time shifts can be used for each RTW period or spike resulting in such a new process, which cannot reversed by a simple operation. However a random time shift at each RTW or spike events is itself a random noise generator thus, from a hardware point of view nothing has been gained.

## 6. Conclusion

While short (shorter-than-correlation-time) deterministic time shifts [10] have already been proposed for a better identification of noise bits, the LTC time shifts described in this Letter have different purpose: expanding the logic hyperspace. In practical applications fixed random time shifts for each logic value would combine the advantages of both time-shifting methods.

1. L.B. Kish, "Noise-based logic: Binary, multi-valued, or fuzzy, with optional superposition of logic states", *Phys. Lett. A* **373** (2009) 911–918. http://arxiv.org/abs/0808.3162 .
2. J. Mullins, "Breaking the noise barrier", *New Scientist*, issue 2780 (29 September 2010); http://www.newscientist.com/article/mg20827801.500-breaking-the-noise-barrier. html?full=true.
3. S.M. Bezrukov, L.B. Kish, "Deterministic multivalued logic scheme for information processing and routing in the brain", *Phys. Lett. A* **373** (2009) 2338-2342. http://arxiv.org/abs/0902.2033 .
4. Z. Gingl, S. Khatri, L.B. Kish, "Towards brain-inspired computing", *Fluct. Noise Lett.* **9** (2010) 403–412. http://arxiv.org/abs/1003.3932 .
5. L.B. Kish, S. Khatri, F. Peper, "Instantaneous noise-based logic", *Fluct. Noise Lett.* **9** (2010) 323-330. http://arxiv.org/abs/1004.2652 .
6. F. Peper, L.B. Kish, "Instantaneous, non-squeezed, noise-based logic", *Fluct. Noise Lett.* **10** (June 2011), open access: http://www.worldscinet.com/fnl/00/0001/open-access/S0219477511000521.pdf
7. L.B. Kish, S. Khatri, S. Sethuraman, "Noise-based logic hyperspace with the superposition of $2^N$ states in a single wire", *Phys. Lett. A* **373** (2009) 1928–1934. http://arxiv.org/abs/0901.3947 .
8. H. Wen, L.B. Kish, A. Klappenecker, "Complex noise-bits and large-scale instantaneous parallel operations with low complexity", *Fluct. Noise Lett.* **12** (2013) 1350002. http://vixra.org/abs/1208.0226 .

9. H. Wen, L.B. Kish, " Noise based logic: why noise? A comparative study of the necessity of randomness out of orthogonality", *Fluct. Noise Lett.* **11** (2012) 1250021. http://arxiv.org/abs/1204.2545 .
10. H. Wen, L.B. Kish, A. Klappenecker, F. Peper, "New noise-based logic representations to avoid some problems with time complexity", *Fluct. Noise Lett.* **11** (2012), 1250003. http://arxiv.org/abs/1111.3859 .
11. L.B. Kish, S. Khatri, T. Horvath, "Computation using Noise-based Logic: Efficient String Verification over a Slow Communication Channel", *Eur. J. Phys. B*. **79** (2011) 85-90. http://arxiv.org/abs/1005.1560 .